# On the effect of surface recombination in thin film solar cells, light emitting diodes and photodetectors

Oskar J. Sandberg, and Ardalan Armin

Department of Physics, Swansea University, Singleton Park, Swansea SA2 8PP Wales, United Kingdom

Email: ardalan.armin@swansea.ac.uk; o.j.sandberg@swansea.ac.uk

**Abstract**

Radiative and non-radiative charge carrier recombination in thin-film diodes plays a key role in determining the efficiency of electronic devices made of next generation semiconductors such as organic, perovskite and nanocrystals. In this work, we show that lowering the bulk recombination does not necessarily result in enhanced performance metrics of electronic devices. From the perspective of charge carrier extraction and injection, the radiative limit of the open-circuit voltage of solar cells, noise current of photodetectors and lasing threshold of injection lasers cannot be improved if the contacts are not perfectly selective. A numerical drift-diffusion model is used to investigate the interplay between bulk recombination and surface recombination of minority carriers at the contacts in bipolar thin diode devices based on low-mobility semiconductors. The surface recombination becomes prominent in case of reduced bulk recombination strengths when non-selective contacts, i. e. contacts that are either metallic or have imperfect charge-selective interlayer, are employed. Finally, we derive analytical approximations for the case when diffusion-limited surface recombination of minority carriers at Ohmic contacts dominates the dark current. These results indicate that having perfectly selective contacts becomes crucial in systems with suppressed bulk recombination – a challenging requirement for future state-of-the-art thin-film solar cells, light-emitting devices and photodetectors made of next generation semiconductors.

## 1. Introduction

Sandwich-type thin-film diode devices based on organic semiconductors and perovskites exhibit great promise for future electronics and energy technology [1, 2, 3, 4]. Of particular interest is bipolar diodes which are essential for a multitude of diode applications, the most important being light-emitting diodes, photodetectors, and solar cells. A bipolar diode consists of an active semiconductor layer which is able to sustain both electron and hole transport. In the simplest thin-film device structure, the active layer, composed of intrinsic semiconductor(s), is sandwiched between two electrodes – a hole-injecting anode and an electron-injecting cathode. Depending on the application, electrons are then either to be injected or collected at the cathode, whereas holes are to be injected or collected at the anode.

Under low-voltage operation, the diode current density generally takes the form

$$J(V) = -J_{\text{ph}}(V) + J_0(V)\left[\exp\left(\frac{qV}{kT}\right) - 1\right] + \frac{V}{R_{\text{sh}}} \quad (1)$$

where $V = V_{\text{ext}} - JR_s$ is the applied voltage over the diode element, $V_{\text{ext}}$ is the external applied voltage, $R_s$ the series resistance of the external circuit and $R_{\text{sh}}$ is the shunt resistance associated with parasitic leakage currents via shorts and defects in the film. Furthermore, $kT/q$ is the thermal voltage and $J_{\text{ph}}$ is the additional photocurrent induced by exposing the diode to external light. Finally, $J_0$ is the so called dark saturation current density. This parameter is highly dependent on the dominating charge carrier recombination mechanism [5, 6, 7] and plays a critical role in the determination of the open-circuit voltage of solar cells [8] and the noise equivalence power of photodetectors [9].

The recombination of charge carriers is one of the most important mechanisms taking place in these diode devices. The recombination can be either radiative or non-radiative. While recombination is in general a loss mechanism in photodetectors and solar cells, radiative recombination is vital for the operation of light-emitting diodes. Moreover, while radiative recombination is unavoidable, non-radiative recombination has commonly been regarded as a parasitic loss mechanism. Non-radiative recombination mainly lowers the quantum efficiency in light-emitting diodes, increases the overall open-circuit voltage loss in solar cells and give rise to increased noise currents in photodetectors.

Another recombination channel, which is also generally associated with non-radiative recombination, is surface recombination at the contacts [10, 11, 12, 13, 14]. In this context, surface recombination refers to the unintentional collection of charge carriers at the "wrong" electrodes, that is, electrons at the anode and holes at the cathode. To avoid this type of recombination, charge-selective interlayers are usually employed in-between the active layer and the electrode [15]. The successful operation of these layers relies on efficient transport of majority carriers (holes at the anode, electrons at the cathode), while simultaneously preventing minority carriers (electrons at the anode, hole at the cathode) from being injected and extracted at the contact in question. The use of nearly ideal interlayers ensures that no charge carrier can escape the device without recombining with a counter charge carrier in LEDs [1]. In photodetectors the blocking contacts are meant to suppress the additional passage of carriers through the device under reverse bias to minimize the dark current in order to reduce the noise current. Similarly, in organic solar cells, lack of perfectly selective electrodes results in charge carriers escaping to the wrong electrodes under open-circuit conditions manifesting as a reduced electroluminescence efficiency with respect to what is expected from the bulk alone.

In organic semiconductor devices, and in particular organic solar cells, the presence of surface recombination remains controversial and the influence of surface recombination on the diode characteristics is often omitted. Simultaneously, in the past few years, a few state-of-the-art donor-acceptor systems have shown significantly reduced recombination and high efficiencies [16, 17], yet no evidence for improvement in the radiative limit of the open circuit voltage has been reported. In this work, the impact of surface recombination on the dark saturation current in bipolar diode devices is investigated by means of analytical derivations and numerical simulations. Surface recombination dominates the dark current in systems with non-selective contacts and exhibiting low bulk recombination rate constants. This shows the key importance of selectivity of contacts especially in systems with reduced recombination. On the other hand, if selective contacts, that prevent surface recombination from taking place, are used the current is instead dominated by bulk recombination processes. Finally, the requirements needed to avoid surface recombination at the contacts are discussed.

## 2. Theory

In this work a drift-diffusion model is used to investigate the electrical transport, accounting for charge-carrier generation and recombination processes, space charge effects, and the properties of the contacts [13, 18]. The charge transport of electrons and holes is described by the charge carrier continuity equations, with the electron and hole current densities $J_n(x)$ and $J_p(x)$, respectively, assumed to follow the drift-diffusion relations for electrons and holes [5, 19]. For electrons, in the dark, the steady-state (time-independent) continuity equation reads

$$\frac{1}{q}\frac{dJ_n}{dx} = \mathcal{R} \qquad (2)$$

where $\mathcal{R}$ is the net bulk recombination rate of charge carriers. Moreover, the classical Einstein relation is assumed to be valid for free charge carriers [20]; subsequently, the electron and hole density $n$ and $p$, respectively, are given by

$$n = N_c \exp\left(\frac{E_{Fn}-E_c}{kT}\right) \quad (3)$$

$$p = N_v \exp\left(\frac{E_v-E_{Fp}}{kT}\right) \quad (4)$$

where $E_{Fn}$ and $E_{Fp}$ are the electron and hole quasi-Fermi levels, respectively. Here, $E_c$ is the (effective) conduction level edge (for the electron transport states), $E_v$ is the (effective) valence level edge (for hole transport states), with $N_c$ and $N_v$ being the associated density of electron and hole transport states, respectively. Note that at thermal equilibrium ($E_{Fn} = E_{Fp}$), the product $np$ is constant and given by the law of mass action $n_i^2 = N_c N_v \exp(-E_g/kT)$, with $E_g = E_c - E_v$ being the electrical energy level gap.

Based on the charge carrier continuity relations, the total current density $J = J_n(x) + J_p(x)$ flowing through the device *in the dark* can be written as

$$J = q \int_0^d \mathcal{R}(x)dx + J_n(0) + J_p(d) = J_R + J_S \quad (5)$$

where $J_R = q \int_0^d \mathcal{R}(x)dx$ is the net bulk recombination current and $J_S = J_n(0) + J_p(d)$ is the net surface recombination current. In forward bias, the total dark current density is thus determined by the net rate of charge carriers recombining within the bulk and the net rate of electrons and holes being collected at the anode and the cathode, respectively. A schematic energy level diagram is shown in Figure 1.

## 2.1. Bulk recombination processes

In general, the bulk recombination rate of charge carriers takes the form

$$R = \beta_0 np \quad (6)$$

with $\beta_0$ being the associated recombination coefficient. Free charge carriers can recombine with each other either directly by band-to-band transitions or indirectly by trap-assisted recombination via impurity states in the gap or Auger recombination [5, 21, 22], the latter two being manifested by a carrier-density dependent recombination coefficient: $\beta_0 = \beta_0(n, p)$. In organic semiconductors, the recombination of free charge carriers is usually taking place via several intermediate steps. In organic homo-junctions made of a single material, the encounter of an electron and a hole gives rise to the formation of an exciton, and the recombination coefficient is usually well approximated by the Langevin recombination rate coefficient $\beta_0 = \beta_L$, where [23, 24]

$$\beta_L = \frac{q}{\varepsilon\varepsilon_0}[\mu_n + \mu_p] \tag{7}$$

as determined by the sum of charge carrier mobility for electrons and holes ($\mu_n$ is the electron mobility, $\mu_p$ is the hole mobility), and the permittivity of the active layer ($\varepsilon\varepsilon_0$).

In organic bulk heterojunction (BHJ) solar cells, however, a different situation usually prevails. Firstly, the confinement of electrons and holes to separate phases, with electrons in the acceptor and holes in the donor, renders the charge encounter rate morphology-dependent [25, 26, 27, 28, 29]. Secondly, once an encounter between an electron (in the acceptor phase) and a hole (in donor phase) do occur, this results in the formation of an intermediate charge-transfer complex at the donor-acceptor interface [30, 31, 32]. This charge transfer complex may then either ultimately recombine into the ground-state, with a rate-coefficient $k_f$, or dissociate back into free charge carriers, with the rate-coefficient $k_d$. Qualitatively, the steady-state kinetics of the charge-transfer complexes, of density $X$, can be described by $dX/dt = 0 = G_X + \beta_0 np - [k_f + k_d]X$. While the encounter rate of free carries (forming CT states) is given by $\beta_0 np$, their net recombination rate to the ground state, after accounting for the back-dissociation of the CT states to separated charge carriers, is given by $\mathcal{R} = \beta_0 np - k_d X$ [33, 34, 35, 36]. Here, $G_X$ is the generation rate of charge transfer complexes from photons mediated by excitons. For simplicity we assume that the quenching yield of the excitons is nearly perfect.

From the condition that the net recombination rate of charge carriers is zero ($\mathcal{R} = 0$) at thermal equilibrium ($np = n_i^2$), it follows that

$$\mathcal{R} = \beta[np - n_i^2] \tag{8}$$

where $\beta = k_f \beta_0 / (k_f + k_d)$ is the reduced recombination with respect to the encounter rate constant $\beta_0$. In Eq. (8), the term $\beta n_i^2$ corresponds to the thermal generation rate of free charge carriers and is given by dissociation probability of the CT states ($P$):

$$\beta n_i^2 = PG_{X,th} = \frac{k_d}{k_f + k_d} G_{X,th} \tag{9}$$

with $G_{X,th}$ being the thermal generation rate of charge transfer complexes. Note that $k_d = (\beta_0 n_i^2/N_X)\exp(E_{ct}/kT)$ and $k_f = (G_{X,th}/N_X)\exp(E_{ct}/kT)$, assuming the thermal equilibrium density of (excited) CT states to be given by $X_0 = N_X \exp(-E_{ct}/kT)$, where $N_X$ is the effective density of (excitable) CT states and $E_{ct}$ is the energy of the CT state. Recent studies suggest that the decay rate constant for CT states is composed of both a radiative and a non-radiative component, implying that $k_f = k_{f,r} + k_{f,nr}$ [37].

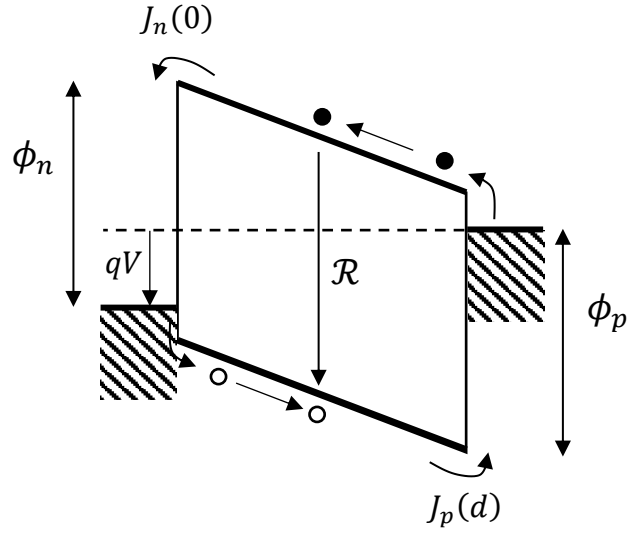

Fig. 1. Schematic energy level diagram in the dark. Upon applying a small positive voltage $V$, over the active layer, electrons are injected at the cathode contact whereas holes are injected at the anode contact. The injected electrons and holes will either recombine inside the active layer with a net bulk recombination rate $\mathcal{R}$, or be collected at the counter electrode. The collection current of electrons at the anode $J_n(0)$ and holes at the cathode $J_p(d)$ is described as surface recombination of minority carriers at the contacts.

## 2.2. Surface recombination of minority carriers at the contacts

The surface recombination current density depends on the charge transport and the charge exchange rates of minority carriers (i.e. electrons at the anode, holes at the cathode) at the contacts. The corresponding electron current density at the anode in organic diode devices is described by [38, 39, 40]

$$J_n(0) = qS_n[n(0) - n_{an}] \tag{10}$$

where $S_n$ is the surface recombination velocity for electrons at anode, whereas $n_{an}$ is the thermal equilibrium density of electrons at the anode given by

$$n_{an} = N_c \exp\left(-\frac{\phi_n}{kT}\right) \tag{11}$$

with $\phi_n = E_c(0) - E_{F,an}$ being the corresponding injection barrier for electrons at the anode.

A contact that is non-selective with respect to charge carrier extraction is generally represented by high values for the surface recombination velocities, typically on the order of $10^5$-$10^6$ cm/s. This is usually the case at contact surfaces with conductive metal electrodes or highly defective interlayers [13, 39, 21]. In low-mobility systems, a high surface recombination velocity is manifested by the surface recombination rate being limited by the bulk transport and not by the kinetics at the contact itself, virtually behaving as $S_n \to \infty$ [5, 13]. For a perfectly selective anode contact, in turn, no electrons are able to pass through the contact (i.e. $J_n(0) = 0$), corresponding to an electron surface recombination velocity $S_n = 0$ at the anode [39]. A selective contact is usually achieved by inserting a minority-carrier-blocking (but majority-carrier-conducting) interlayer between the metal-like electrode and the active layer. However, even in case of a nominally selective contact, the presence of impurity-induced gap states (via which the transport of minority carriers can take place) or the possibility for (direct) recombination with majority carriers from the interlayer, might ultimately render the contact non-selective [15]. These are rather intrinsic limitations in realizing perfectly selective contacts in practice.

In order to explicitly relate the surface recombination current Eq. (10) at the anode to bulk properties of the active layer, consider the transport of electrons being collected at the anode. The electron current equation is given by

$$J_n(x) = \mu_n n \frac{dE_c}{dx} + \mu_n kT \frac{dn}{dx} = J_n(0) + J_r(x) \tag{12}$$

where the last equality follows from integrating the electron continuity equation (Eq. (2)) with $J_r(x) \equiv q \int_0^x \mathcal{R}(x')\, dx'$. Eq. (12) may be solved assuming that the following conditions apply: i) the electron injection at the cathode ($x = d$) is ideal with electrons being at thermal equilibrium at the cathode contact: $E_{Fn}(d) = E_{F,cat}$ (related to $n(d)$ via Eq. (3)), ii) the applied voltage $V$ is given by the electrode Fermi level difference, $qV = E_{F,cat} - E_{F,an}$, and iii) the electron density at the anode $n(0)$ is given in accordance with Eq. (10).

Under these conditions, after multiplying Eq. (12) by $\exp([E_c(x) - E_{F,an}]/kT)$, and integrating from $x = 0$ to $x = d$, the net surface recombination current of electrons at the anode is obtained as:

$$J_n(0) = \frac{qS_n n_{an}}{1+S_n/v_n}\left[\exp\left(\frac{qV}{kT}\right) - 1\right] - \frac{qS_n}{1+S_n/v_n}\int_0^d \frac{J_r(x)}{\mu_n kT}\exp\left(\frac{E_c(x)-E_{F,an}-\phi_n}{kT}\right)dx \quad (13)$$

with

$$v_n = \left[\int_0^d \frac{q}{\mu_n kT}\exp\left(\frac{E_c(x)-E_{F,an}-\phi_n}{kT}\right)dx\right]^{-1} \quad (14)$$

being the effective transport velocity for electrons diffusing towards the anode. A completely analogous expression can be found for the surface recombination current density of holes $J_p(d)$ at the cathode.

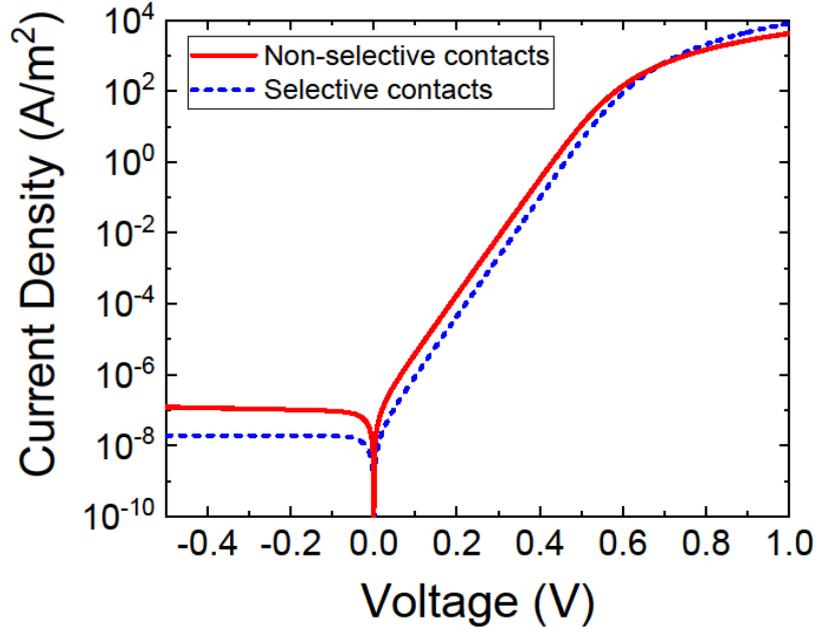

Fig. 2. Simulated current-voltage characteristics in the dark for a device with two non-selective contacts ($S_n = S_p \to \infty$) and two selective contacts ($S_n = S_p \to 0$). The bulk recombination coefficient is assumed to be $\beta = 0.01 \times \beta_L$. Injection barriers of 0.20 eV for majority carriers are assumed. The electrical bandgap is assumed to be $E_g = 1.1$ eV.

## 3. Results

The effect of charge-selective contacts on the dark current density of a bipolar diode is simulated in Figure 2. A constant bulk recombination coefficient of $\beta = 0.01 \times \beta_L$ is assumed. The active layer thickness is 100 nm, and equal electron and hole mobilities of $\mu_n = \mu_p = 3 \times 10^{-4}$ cm²/Vs are assumed. The electrical bandgap is assumed to be $E_g = 1.1$ eV, whereas $N_c = N_v = 10^{21}$ cm$^{-3}$. The contacts are assumed to be ideal in terms of the injection/extraction velocities for the majority carriers, with majority-carrier injection barriers of 0.2 eV at both contacts. The corresponding injection barriers for minority carriers are in this case given by $\phi_n = \phi_p = 0.9$ eV (where $\phi_p = E_{F,cat} - E_v(d)$ is the injection barrier for holes at the cathode). Finally, we assume $1/R_{sh} = 0$ (see Eq. (1)), thus neglecting the influence of parasitic shunts.

From Figure 2, it can be seen that by changing both contacts from non-selective to selective, a decrease in the magnitude of the current is obtained. This decrease is a consequence of removing the surface recombination, leading to a reduced recombination current overall. The total diode current reads

$$J = J_0 \left[ \exp\left(\frac{qV}{kT}\right) - 1 \right] \qquad (15)$$

where $J_0 = J_{R,0} + J_{S,0}$. In Figure 3, the impact of the bulk recombination strength on the dark saturation current $J_0$, defined in accordance with Eq. (15), is simulated for the devices with selective and non-selective contacts. It can be seen that, for non-selective contacts, surface recombination becomes dominant in case of low bulk recombination strengths. Depending on the interplay between the bulk recombination strength $\beta/\beta_L$ and the properties of the contacts, different situations can be identified.

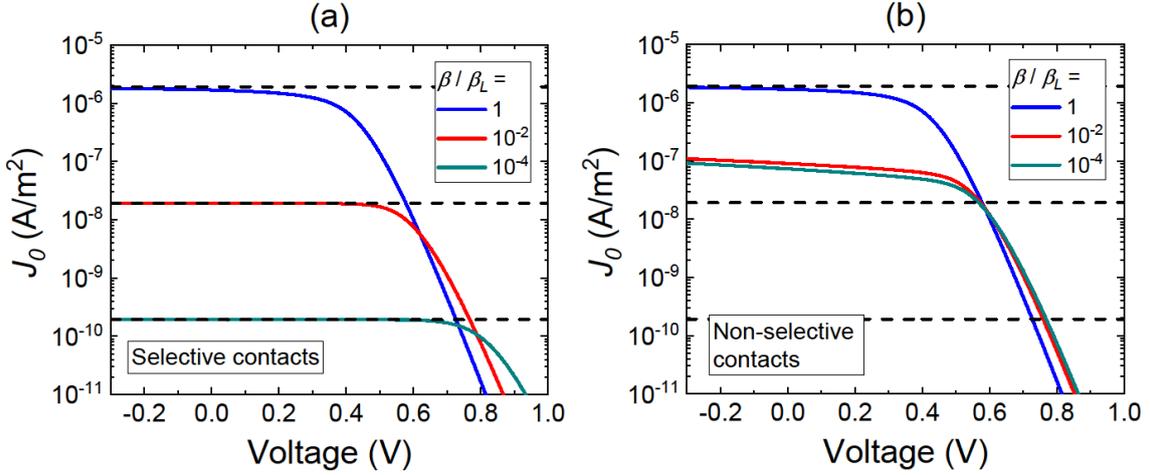

Fig. 3. Simulated dark saturation current densities $J_0(V)$, as a function of the voltage, is shown at different bulk recombination strengths for the case with a) selective contacts and b) non-selective contacts. For comparison, the analytical prediction Eq. (17), expected when bulk recombination dominates the current, has been included as depicted by the dashed lines.

### 3.1. Bulk recombination dominates

In case of strong recombination inside the bulk, or if perfectly selective contacts are used, the current density is dominated by bulk recombination. Under these conditions, the current density under low-voltage operation reduces to

$$J \approx J_R = J_{R,0}\left[\exp\left(\frac{qV}{kT}\right) - 1\right] \qquad (16)$$

Then, under the approximation of constant quasi-Fermi level splitting throughout the active layer, $E_{Fn}(x) - E_{Fp}(x) \approx qV$, and assuming the charge carrier encounter rate to be dominated by a bimolecular process, $J_{R,0}$ is given by

$$J_{R,0} = q\beta n_i^2 d \qquad (17)$$

with $\beta n_i^2$ being the thermal generation rate of separated carriers in the bulk in accordance with Eq. (9). In general, the bulk recombination dominates over surface recombination when $J_{R,0} \gg J_{S,0}$. At high bulk recombination levels, also the density of minority carriers at the contacts are lower, which further tends to suppress the surface recombination. Comparing the analytical approximation Eq. (17) with the simulated dark saturation current densities in Figure 3, it can be seen that bulk recombination dominates (and Eq. (17) is valid) only in case of selective contacts and/or high bulk recombination coefficients.

It should be noted that in case of significant levels of trap-assisted or Auger recombination, the recombination coefficient becomes carrier-density dependent as well: $\beta = \beta(n,p)$. In this case, $\beta$ in Eq. (17) needs to be replaced by its spatial average $\langle \beta \rangle = (1/d) \int_0^d \beta(n,p) dx$ (assuming the quasi-Fermi level splitting remains constant throughout the bulk); this will ultimately result in an additional voltage-dependent $J_{R,0} = J_{R,0}(V)$ [6, 7]. For example, in case of trap-assisted recombination via deep midgap states within the bulk, we expect $J_{R,0}(V) \propto \exp(-qV/2kT)$ [41].

### 3.2. Surface recombination of minority carriers at the contacts dominates

In case of non-selective contacts, on the other hand, a different situation arises in the limit of weak bulk recombination. In this limit the surface recombination (and injection), as manifested by diffusion-limited current transport of minority carriers, dominates over recombination (and generation) processes in the bulk, corresponding to $J_{S,0} \gg J_{R,0}$. The current is then ruled by the surface recombination current,

$$J \approx J_S = J_{S,0} \left[ \exp\left(\frac{qV}{kT}\right) - 1 \right] \tag{18}$$

where

$$J_{S,0} = qv_n N_c \exp\left(-\frac{\phi_n}{kT}\right) + qv_p N_v \exp\left(-\frac{\phi_p}{kT}\right) \tag{19}$$

with the effective diffusion velocity $v_n$ for electrons defined by Eq. (14); an analogous expression applies for the effective diffusion velocity $v_p$ for holes. Eq. (19) assumes that the surface recombination is limited by diffusion of minority carriers to the contact and not by the kinetics at the contact; this is valid as long as $v_{n/p} \ll S_{n/p}$ (here, $S_p$ is the surface recombination velocity for holes at the cathode).

Figure 4 shows the dark saturation current density for the case when bulk recombination is negligible and the contacts are non-selective. Under these conditions, the current is solely dominated by surface recombination of minority carriers at the contacts. This type of situation may arise in case of large minority carrier lifetimes (reduced bulk recombination) giving rise to non-negligible minority carrier densities near the contacts. As expected, the current level is also strongly dependent on the energetic injection barrier for minority carriers at the contacts, increasing exponentially with decreasing injection barrier. In other words, the larger one of the injection barriers for majority carriers is, i.e. the more un-optimized the energy levels at the contact are, the larger will the surface recombination at said contact be. It can be seen however that, depending on the injection barrier, two different principal regimes can be identified. These two regimes are considered next.

### 3.2.1. Surface recombination at nearly neutral contacts

For low-voltage operation, the (diffusion-limited) minority carrier current is dominated by the potential energy landscape close to the contact (but inside the active layer). If the hole density at the anode contact is not too large, the influence of the hole-induced space charge on the prevailing electric field may be neglected. In this case, the energy level bending at the anode contact is small and the contact may be regarded as a "neutral" contact. Under these conditions, the potential energy level for electrons can be approximated as $E_c(x) \approx E_c(0) + qF_0 x$, where $F_0 = [V - V_{bi}]/d$ is the electric field in the active layer and $V_{bi}$ is the built-in voltage. The associated diffusion velocity Eq. (14) for electrons can then be readily evaluated as

$$v_n = \frac{\mu_n [V_{bi} - V]}{d} \left[ 1 - \exp\left(\frac{q[V - V_{bi}]}{kT}\right) \right]^{-1} \tag{20}$$

For voltages well below $V < V_{bi}$, the diffusion velocity is thus well approximated by the electron drift velocity $v_n \approx \mu_n |F_0|$ at the anode.

Comparing the analytical approximation Eq. (18) and (19), in conjunction with Eq. (20), with the simulated currents in Figure 4, it can be seen that Eq. (20) provides an excellent approximation for the diffusion velocity at the lowest minority-carrier injection barriers $\phi_n$ and $\phi_p$. At these $\phi_n$ and $\phi_p$, the injection barriers for majority carriers is large enough ($\geq 0.2$ eV) for the space charge from the majority carriers at the contacts to be negligible. It should be noted that for low voltages ($0 < V < V_{bi}$) the surface recombination current of electrons is only limited by the electric field close to the anode region. The effect of possible energy-level bending at the counter electrode (i.e. the cathode) is in this case to effectively decrease the built-in voltage $V_{bi}$ inside the active layer [42], leaving Eq. (20) otherwise unchanged (for $V < V_{bi}$). Under these conditions, $V_{bi}$ should be considered an effective quantity which also accounts for the potential drop caused by energy-level bending at the counter electrode.

We noted that Eq. (18) and (19), with the effective diffusion velocity given by Eq. (20), is identical to previous expressions obtained by others to describe diffusion-limited currents in unipolar diodes [42, 43, 44]. Furthermore, we have omitted the image-force lowering of the minority-carrier injection barrier at the contacts [5, 45, 46]. Accounting for this effect generally leads to an increase of $v_n$ (relative to Eq. (20)) by a factor $[4kT/(\pi q|F_0|r_c)]^{1/4} \times \exp(\sqrt{-qF_0 r_c/kT})$ at high reverse bias ($V < 0$) [47], where $r_c \equiv \frac{q^2}{4\pi\varepsilon\varepsilon_0 kT}$. At (nearly) Ohmic contacts, however, the image charge effects are commonly assumed to be screened by the injected charge carriers, and the barrier-lowering effect may be neglected [24].

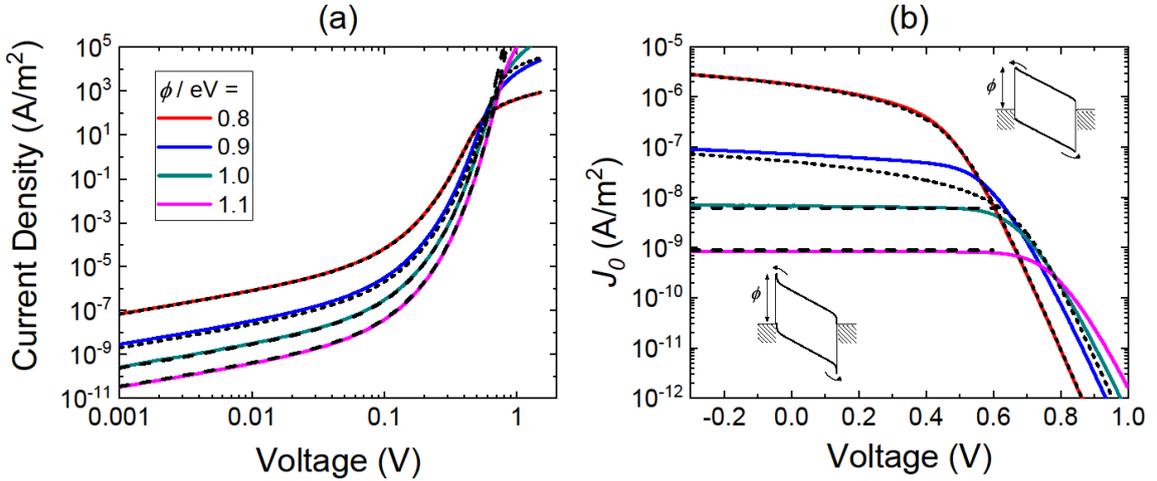

Fig. 4. (a) Simulated dark current-voltage characteristics is shown for the case with *negligible* bulk recombination and non-selective contacts for different injection barriers $\phi = \phi_n = \phi_p$ for minority carriers at the contacts. In (b), the corresponding dark saturation current densities $J_0(V)$, as a function of the voltage, of the current from (a) is shown. For comparison, the analytical approximation Eq. (18) and Eq. (19) have been included and depicted by the short-dashed line for the case with minority-carrier diffusion at nearly neutral contacts (Eq. (20)) and by the long-dashed lines for the case with minority-carrier diffusion at a contact with strong energy level bending (Eq. (23)).

*3.2.2. Surface recombination of minority carriers at heavily injecting contacts*

As the minority-carrier injection barriers $\phi_n$ and $\phi_p$ are increased, however, the injection level of majority carriers increases drastically, and the energy-level bending at the contacts become important. Under these conditions, the contact region is highly conducting for majority carriers, the contact in this case corresponding to an Ohmic contact. For the case with a hole-Ohmic anode contact, the energy level bending is caused by the accumulated space charge of injected holes at this contact. Since the region close to this contact, by definition, is not limiting the current transport for majority carriers (holes), the majority carrier density is very closely at thermal equilibrium within this region (during low-voltage operation).

The Poisson equation in the vicinity of the Ohmic anode contact is given by

$$\frac{d^2 E_c}{dx^2} = \frac{d^2 E_v}{dx^2} \approx \frac{qp(x)}{\varepsilon \varepsilon_0} \tag{21}$$

Approximating the hole density by Eq. (4) and taking $E_{Fp} \approx E_{F,an}$ within this region, the Poisson equation can be solved analytically. In the hole-dominated region close to the anode we find

$$E_c(x) \approx E_c(0) - 2kT \ln\left(1 + \frac{x}{L_{an}}\right) \tag{22}$$

with the characteristic screening length given by $L_{an} = \sqrt{2\varepsilon\varepsilon_0 kT/(q^2 p_{an})}$, where $p_{an}$ is the (thermal equilibrium) hole density at the anode contact. The diffusion velocity of electrons, diffusing against the (hole-induced) upward energy-level bending is then obtained as

$$v_n = \frac{\mu_n N_c kT}{L_{an}} \times \eta_{DOS} \tag{23}$$

where we have introduced an additional correction factor $\eta_{DOS}$ to account for the density-of-state (DOS) filling effects for majority carriers; under non-degenerate conditions, when Eq. (22) applies and the details of the DOS is unimportant, we have $\eta_{DOS} = 1$.

Upon comparing the analytical approximation Eq. (18) and Eq. (19), but with $v_n$ given by Eq. (23), with the simulations in Figure 4 an excellent agreement is indeed obtained for the cases with the highest minority-carrier injection barriers when the energy-level bending induced by the majority carriers becomes significant (see lower inset in Figure 4(b)). An interesting feature of the surface recombination current in this regime (Eq. (23)) is that it is independent of the active layer thickness. This is a direct consequence of the surface recombination being limited by diffusion in the thin space charge region close to the Ohmic contact.

### 3.2.3. DOS filling effects at the contact

Eq. (22) is strictly valid only for non-degenerate conditions, $p_{an} \ll N_v$, when Boltzmann statistics applies. Under degenerate conditions, however, when $p_{an} \sim N_v$, DOS filling effects for the injected holes at the anode contact becomes important [48]. Close to the anode, the hole density is then obtained by integrating the product of Fermi-Dirac distribution and density of states $g_p(E, x)$ over energy

$$p(x) = \int_{-\infty}^{\infty} \frac{g_p(E,x)}{1+\exp\left(-\frac{E-E_{F,an}}{kT}\right)} dE \tag{24}$$

assuming $E_{Fp} \approx E_{F,an}$ in the accumulation region.

Figure 5 shows the effect of DOS filling ($\eta_{DOS}$) on the surface recombination current for different disorder parameters $\sigma$, assuming a Gaussian DOS of the form $g_p(E,x) = \left(N_v/\sqrt{2\pi\sigma^2}\right)\exp\left(-[E-E_{v,0}(x)]^2/2\sigma^2\right)$. The degree of DOS occupation of holes at the anode contact is strongly dependent on the energetic offset $\Delta = E_{F,an} - E_{v,0}(0)$ between the anode Fermi level and the Gaussian DOS center for the hole transport states at the anode. The effect of DOS filling is to extend the energy-level bending region further into the bulk, leading to an increase of the effective diffusion velocity of electrons at the anode (by the factor $\eta_{DOS} > 1$). Furthermore, the impact of DOS filling increases with increasing energetic disorder $\sigma$.

For large disorder parameters $\sigma \gg kT$ and/or significant DOS filling for holes (at the anode contact), the enhancement factor $\eta_{DOS}$ of the diffusion velocity of electrons diffusing against the hole-induced energy level bending at the anode contact can be approximated by

$$\eta_{DOS} \approx \sqrt{\frac{4\sigma}{kT}\left[\frac{\exp\left(-\frac{\Delta^2}{2\sigma^2}\right)}{c\sqrt{2\pi}} - \frac{\Delta}{\sigma}\right]} \tag{25}$$

if $\eta_{DOS} > 1$, where $c = \frac{1}{2}\text{erfc}\left(\frac{\Delta}{\sigma\sqrt{2}}\right)$ corresponds to the DOS occupation of holes at the anode contact in the degenerate limit ($T \to 0$ K). In the derivation of Eq. (25), the following three approximations were made: i) $p(x) \approx \int_{E_{F,an}}^{\infty} g_p(E,x)\, dE = \frac{N_v}{2}\text{erfc}\left(\frac{[E_{F,an}-E_{v,0}(x)]}{\sigma\sqrt{2}}\right)$, ii) the magnitude of the electric field and the hole density is much larger at the anode than in the bulk, and iii) that $v_n \approx (\mu_n/q)|dE_{v,0}(0)/dx|$. Eq. (25) is then readily obtained after multiplying both sides in Eq. (21) by $dE_{v,0}/dx$ and integrating.

It should be stressed that a constant mobility, independent of carrier concentration and the electric field, has been assumed for the electrons. This can be considered a valid approximation at low-voltage operation. Subsequently, the DOS filling (for the majority carriers at the contact) only affects the minority carrier transport indirectly via Eq. (21). We note, however, that if the DOS is assumed to be Gaussian for electrons as well, the effective conduction level edge for electrons (following non-degenerate statistics close to the anode contact) will then be given by $E_c = E_{c,0} - \sigma^2/2kT$, where $E_{c,0}$ is the corresponding DOS center for the electron transport states [49].

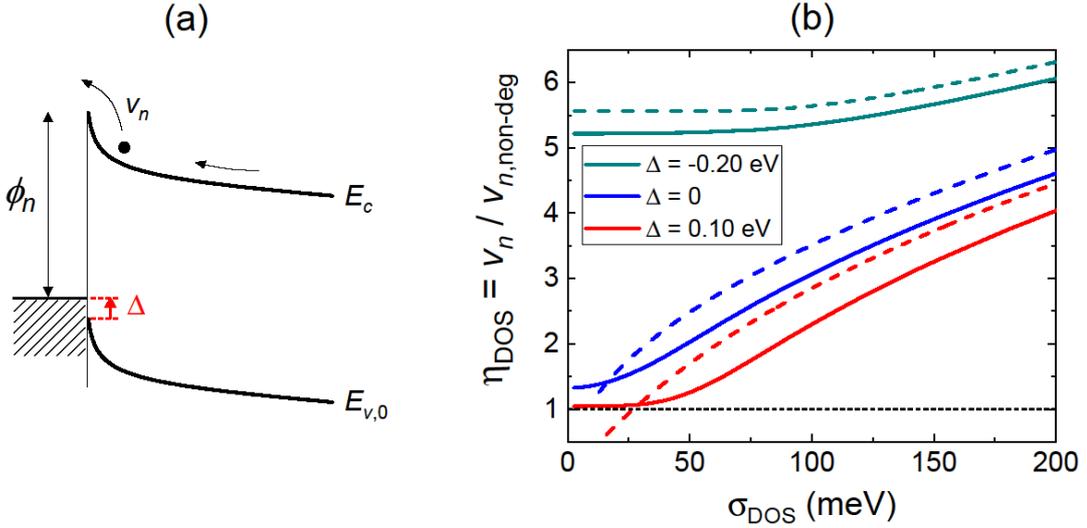

Fig. 5. In a) the energy level diagram close to the hole-Ohmic anode contact is shown for the situation when a considerable hole-induced energy level bending is present. The electron collection is limited by diffusion against the energy level bending at the hole contact. In b) the corresponding enhancement factor $\eta_{DOS}$ of the associated electron diffusion velocity, as induced by DOS filling effects of holes at the Ohmic contact, is simulated (solid lines). The analytical approximation Eq. (25) is depicted by the dashed lines.

## 4. Discussion

The condition for surface recombination not to be dominating the current is given by $J_{S,0} \ll J_{R,0}$. If the surface recombination is dominated by the collection of electrons at the anode, assumed to be non-selective in terms of carrier collection ($v_n \ll S_n$), then this condition can be rewritten as

$$\beta \gg \frac{v_n}{N_v d} \exp\left(\frac{E_g - \phi_n}{kT}\right) \tag{26}$$

where $v_n \sim \mu_n |F(0)|$ for $V < V_{bi}$. To minimize the surface recombination, the energy level offset ($E_g - \phi_n$) for majority carriers at the contact needs to be as small as possible. However, as evident from Fig. 3, even at relatively low injection barriers of $E_g - \phi_n = 0.20$ eV, a significant surface recombination of minority carriers is still present in the device, emphasizing the need for selective contacts. This is of particular importance in devices with low bulk recombination rates.

### 4.1. The open circuit voltage of solar cell

For over a decade, most high-efficiency organic BHJ systems have been found to exhibit recombination coefficient close to Langevin, $\beta = \zeta \beta_L$ with reduction factors $\zeta$ around 0.01-0.1 [50, 51, 52]. It is therefore not surprising that the surface recombination in such systems (majority of organic donor/acceptor blends) has often been ignored since it plays an insignificant role. In more recent years, there have also been a few reports of BHJ systems with high efficiencies exhibiting significantly suppressed bulk recombination (relative to Langevin) [16, 17], however, evidence for reduced radiative losses in the $V_{oc}$ is still lacking. From the above discussion about the crucial role of surface recombination in systems with small bulk recombination, it can be understood that surface recombination is a limiting factor in determining the $V_{oc}$ of both non-Langevin BHJ systems and perovskite solar cells. It is also important to note that surface recombination may complicate the determination of non-radiative losses via charge transfer states in organic solar cells [37]. As seen from Fig. 3(b), the surface recombination in non-Langevin type systems ($\zeta \ll 1$) increases $J_0$ by several orders of magnitude and this may exceed the non-radiative limit of $J_0$ in the bulk. In this case, the surface recombination in systems with significantly small bulk recombination may easily be confused by non-radiative environmental-assisted recombination.

## 4.2. Light-emitting diodes and lasers

In light-emitting diodes it is well established that charge-blocking layers are required in order to "confine" the carriers in the device [1]. In LED nomenclature the term surface recombination is sometimes omitted but the concept is well known in that area of research. One striking aspect of surface recombination is the limitations it sets in achieving organic injection lasers. A hypothetical injection organic laser requires extremely high carrier density in order to achieve population inversion at which the quasi-Fermi level splitting exceeds the bandgap. Such conditions may be challenging to achieve with typical Langevin type materials in which electrons and holes recombine as soon as they meet in space. In non-Langevin systems, in turn, the lifetime is much longer and electrons and holes can exist in the bulk, forming a plasma. However, at small recombination rates in the bulk, the surface recombination becomes dominant – pinning the quasi-Fermi levels to the contacts at high injection levels and making the population inversion nearly impossible. Even a very small surface recombination velocity would result in rather large surface recombination current are high injection levels. An important obstacle in front of organic injection laser is therefore not only a highly luminescent semiconductor with small bulk recombination but also extremely selective contacts which are difficult to achieve.

## 4.3. Photodetectors and surface recombination

The most important performance metric for photodetectors is their noise equivalent power; the power at which the signal equals the noise, i.e. when the photocurrent has the same magnitude as the noise current. The noise current has multiple components including frequency-dependent flicker and generation/recombination noise and frequency-independent thermal and shot noise [3, 9]. In the ideal case, when the shot noise is dominant in reverse bias, the noise power is given by

$$\langle j^2 \rangle = 2qJ_0 \Delta f$$

where $\Delta f$ is the electrical bandwidth and $J_0$ is the (average) dark saturation current, as before. Ideally, $J_0$ is only determined by radiative bulk process. Taking non-radiative bulk recombination of charge transfer states into account, $J_0$ is increased by a factor of $\text{EQE}_{\text{EL}}^{-1}$, where $\text{EQE}_{\text{EL}}$ is the external quantum efficiency for electroluminescence [7]. Unfortunately, significant non-radiative recombination appears to be an intrinsic limitation of organic BHJ systems and increases with decreasing the bandgap [37], required for near-infra red (NIR) photodetectors [53]. As such, organic semiconductors do not seem to be entirely suitable for sensitive NIR detection with $EQE_{EL}$ values less than $10^{-6}$ in narrow gap systems [54]. However, as show in Fig. 3(a), suppressing the bulk recombination can result in significant reduction of $J_0$ provided the contacts are as selective as possible. Again, this is a challenging requirement but presents a hope for NIR detection using organic semiconductors.

## 5. Conclusions

In summary, the surface recombination of minority carriers at non-selective contacts becomes important in bipolar thin-film diodes with low bulk recombination rates and/or high mobilities. Under low-voltage operation, the surface recombination at non-selective contacts in thin-film diodes based on low-mobility semiconductors, is governed by diffusion of minority carriers towards the contacts, and increases with increasing carrier mobility and decreasing injection barrier for minority carriers. In case of Ohmic contacts, the surface recombination of minority carriers is limited by diffusion against the energy-level bending at the contact. Since surface recombination give rise to increased (non-radiative) recombination and dark current levels, this ultimately also results in increased open-circuit voltage losses of organic solar cells and noise current levels in thin-film photodetectors. In thin-film devices with low bulk recombination rates and/or un-optimized energy levels at the contacts, the use of selective contacts is therefore of paramount importance.

**Acknowledgements**

The work was supported by the Sêr Cymru Program through the European Regional Development Fund, Welsh European Funding Office and Swansea University strategic initiative in Sustainable Advanced Materials.

[1]  R. Meerheim, B. Lüssem, and Karl Leo, Proc. IEEE 97, 1606 (2009).

[2]  L. Dou, J. You, Z. Hong, Z. Xu, G. Li, R. A. Street, and Y. Yang, Adv. Mater. 25, 6642 (2013).

[3] R. D. Jansen-van Vuuren, A. Armin, A. K. Pandey, P. L. Burn, and P. Meredith, Adv. Mater. 28, 4766 (2016).

[4] C. Deibel and V. Dyakonov, Rep. Prog. Phys. 73, 096401 (2010).

[5] S. M. Sze, Physics of Semiconductor Devices, 3rd ed. (Wiley & Sons, New York, 1981).

[6] T. Kirchartz, B. E. Pieters, J. Kirkpatrick, U. Rau, and J. Nelson, Phys. Rev. B 83, 115209 (2011).

[7] K. Tvingstedt and C. Deibel, Adv. Energy Mater. 6, 1502230 (2016).

[8] J. Yao, T. Kirchartz, M. S. Vezie, M. A. Faist, W. Gong, Z. He, H. Wu, J. Troughton, T. Watson, D. Bryant, and J. Nelson, Phys. Rev. Applied 4, 014020 (2015).

[9] Y. Fang, A. Armin, P. Meredith, and J. Huang, Nat. Photon. 13, 1-4 (2019).

[10] Z. Tang, J. Wang, A. Melianas, Y. Wu, R. Kroon, W. Li, W. Ma, M. R. Andersson, Z. Ma, W. Cai, W. Tress, and O. Inganas, J. Mater. Chem. A, 6, 12574 (2018).

[11] A. Wagenpfahl, C. Deibel, and V. Dyakonov, IEEE J. Sel. Top. Quantum Electron. 16, 1759 (2010).

[12] J. Reinhardt, M. Grein, C. Bühler, M. Schubert, and U. Würfel, Adv. Energy Mater. 4, 1400081 (2014).

[13] O. J. Sandberg, A. Sundqvist, M. Nyman and R. Österbacka, Phys. Rev. Appl. 5, 044005 (2016).

[14] I. Zonno, B. Krogmeier, V. Katte, D. Lübke, A. Martinez-Otero, and T. Kirchartz, Appl. Phys. Lett. 109, 183301 (2016).

[15] E. L. Ratcliff, B. Zacher, and N. R. Armstrong, J. Phys. Chem. Lett. 2, 1337 (2011).

[16] A. Armin, J. Subbiah, M. Stolterfoht, S. Shoaee, Z. Xiao, S. Lu, D. J. Jones, P. Meredith, Adv. Energy Mater. 6, 1600939 (2016).

[17] Y. Jin, Z. Chen, S. Dong, N. Zheng, L. Ying, X.-F. Jiang, F. Liu, F. Huang, and Y. Cao, Adv. Mater. 28, 9811–9818 (2016).

[18] O. J. Sandberg, M. Nyman and R. Österbacka, Phys. Rev. Applied 1, 024003 (2014).

[19] S. Selberherr, Analysis and Simulation of Semiconductor Devices (Springer-Verlag, Wien, 1984).

[20] G. A. H. Wetzelaer, L. J. A. Koster, and P.W. M. Blom, Phys. Rev. Lett. 107, 066605 (2011).

[21] P. Würfel, Physics of Solar Cells, 2nd ed. (Wiley-VCH, Weinheim, Germany, 2009).

[22] L. M. Pazos-Outón, T. P. Xiao, and E. Yablonovitch, J. Phys. Chem. Lett. 9, 1703 (2018).

[23] M. P. Langevin, Ann. Chim. Phys. 28, 433 (1903).

[24] R. Coehoorn and P. A. Bobbert, Phys. Status Solidi A 209, 12 (2012).

[25] C. Groves and N. C. Greenham, Phys. Rev. B 78, 155205 (2008).

[26] L. J. A. Koster, V. D. Mihailetchi, P. W. M. Blom, Appl. Phys. Lett. 88, 052104 (2006).

[27] M. C. Heiber, C. Baumbach, V. Dyakonov, and C. Deibel, Phys. Rev. Lett. 114, 136602 (2015).

[28] A. Pivrikas, N. S. Sariciftci, G. Juška and R. Österbacka, Prog. Photovolt: Res. Appl. 15, 677 (2007).

[29] M. Nyman, O. J. Sandberg, and R. Österbacka, Adv. Energy Mater. 5, 1400890 (2015).

[30] K. Vandewal, K. Tvingstedt, A. Gadisa, O. Inganäs, and J. V. Manca, Phys. Rev. B 81, 125204 (2010).

[31] C. Deibel, T. Strobel, and V. Dyakonov, Adv. Mater. 22, 4097 (2010).

[32] D. H. K. Murthy, A. Melianas, Z. Tang, G. Juska, K. Arlauskas, F. Zhang, L. D. A. Siebbeles, O. Inganäs, and T. J. Savenije, Adv. Funct. Mater. 23, 4262 (2013).

[33] C. L. Braun, J. Chem. Phys. 80, 4157 (1984).

[34] L. J. A. Koster, E. C. P. Smits, V. D. Mihailetchi, and P. W. M. Blom, Phys. Rev. B 72, 085205 (2005).

[35] W. Tress, K. Leo, and M. Riede, Phys. Rev. B 85, 155201 (2012).

[36] A. Armin, J. R. Durrant, and S. Shoaee, J. Phys. Chem. C 121, 13969 (2017).

[37] J. Benduhn, K. Tvingstedt, F. Piersimoni, S. Ullbrich, Y. Fan, M. Tropiano, K. A. McGarry, O. Zeika, M. K. Riede, C. J. Douglas, S. Barlow, S. R. Marder, D. Neher, D. Spoltore, and K. Vandewal, Nature Energy 2, 17053 (2017).

[38] P. S. Davids, I. H. Campbell, D. L. Smith, J. Appl. Phys. 82, 6319 (1997).

[39] T. Kirchartz, B. E. Pieters, K. Taretto, and U. Rau, Phys. Rev. B 80, 035334 (2009).

[40] O. J. Sandberg, S. Sandén, A. Sundqvist, J.-H. Smått, and R. Österbacka, Phys. Rev. Lett. 118, 076601 (2017).

[41] C. T. Sah, R. N. Noyce, W. Shockley, Proc. IRE, 1228 (1957).

[42] P. de Bruyn, A. H. P. van Rest, G. A. H. Wetzelaer, D. M. de Leeuw, and P. W. M. Blom, Phys. Rev. Lett. 111, 186801 (2013).

[43] H. Nguyen, S. Scheinert, S. Berleb, W. Brutting, and G. Paasch, Org. Electron. 2, 105 (2001).

[44] P. Kumar, S. C. Jain, V. Kumar, S. Chand, and R. P. Tandon, J. Appl. Phys. 105, 104507 (2009).

[45] J. C. Scott and G. G. Malliaras, Chem. Phys. Lett. 299, 115 (1999).

[46] Yu. A. Genenko, S. V. Yampolskii, C. Melzer, K. Stegmaier, and H. von Seggern, Phys. Rev. B 81, 125310 (2010).

[47] P. R. Emtage and J. J. O'dwyer, Phys. Rev. Lett. 16, 356 (1966).

[48] I. Lange, J. C. Blakesley, J. Frisch, A. Vollmer, N. Koch, and D. Neher, Phys. Rev. Lett. 106, 216402 (2011).

[49] J. C. Blakesley and N. C. Greenham, J. Appl. Phys. 106, 034507 (2009).

[50] M. Stolterfoht, A. Armin, B. Philippa, and D. Neher, J. Phys. Chem. Lett. 7, 4716 (2016).

[51] T. M. Clarke, C. Lungenschmied, J. Peet, N. Drolet, and A. J. Mozer, J. Phys. Chem. C 119, 7016 (2015).

[52] D. Bartesaghi, I. del Carmen Pérez, J. Kniepert, S. Roland, M. Turbiez, D. Neher, and L. J. A. Koster, Nat. Commun. 6, 7083 (2015).

[53] A. Armin, R. D. Jansen-van Vuuren, N. Kopidakis, P. L. Burn, and P. Meredith, Nat. Commun. 6, 6343 (2015).

[54] M. A. Green, Prog. Photovolt. 20, 472 (2012).